\documentstyle[preprint,aps,prb]{revtex}

\tightenlines
\begin{document}
\draft
\preprint{S/F/S, Feb. 1999}

\title{ Theory of the Josephson effect in superconductor
/ ferromagnet / superconductor
junctions }
\author{Y. Tanaka}

\address{Department of Applied Physics, Nagoya University, Nagoya,
464-8603, Japan.}

\author{S. Kashiwaya}

\address{Electrotechnical Laboratory, Umezono, Tsukuba, Ibaraki
305-8568, Japan.}

\date{March 1, 1999}
\maketitle
\begin{abstract}
To reveal the influence of
the exchange interaction on the Josephson effect,
the d.c. Josephson current in
unconventional-superconductor /
ferromagnet /  unconventional-superconductor
junctions is studied.
When the  two superconductors have $d$-wave symmetry,
the Josephson current is  drastically suppressed
with the increase of the
exchange interaction.
On the other hand, when the two superconductors
have different parities, $e.g.$ $s$-wave and $p$-wave,
an enhancement of the Josephson current is obtained
with the increment of the exchange interaction
due to the break-down of SU(2) symmetry in the spin space.
\end{abstract}
\par
\newpage
\narrowtext
The Josephson effect in superconductor / ferromagnet /superconductor
($S/F/S$) junctions has been theoretically studied since long time ago
\cite{Buzdin1,Buzdin2}.
The most interesting phenomena obtained in
these works is the oscillating behavior of
the Josephson current under the influence of
the exchange interaction \cite{Buzdin1,Demler}.
In some cases, $\pi$-junction \cite{Bulaev}
is attainable by choosing a proper
thickness of the ferromagnet
and the magnitude of the exchange interaction.
However, all existing theories are intended to
conventional  superconductors
and there has been presented no theory
treating  unconventional
superconductors whose pair potentials have internal phases.
\par
Recent tunneling theories for unconventional superconductors
have revealed
several novel features which are not expected for
the conventional BCS superconductor junctions.
First of all, the formation of the
zero energy states (ZES)
at the surfaces of $d$-wave superconductor
has been predicted theoretically \cite{Hu}.
The appearance and the disappearance
of the ZES at the surfaces of high-$T_{c}$
superconductors
have been  studied
in several tunneling spectroscopy experiments,
and the consistency between theory and experiments
has been checked in details
\cite{Tanaka2,Kash,Alff,Ekin,Wei}.
As for the d.c. Josephson effect in anisotropic singlet
superconductors,
we have presented
a general formula (referred to as TK formula) which fully
takes account of the internal phase of the pair potential
including the ZES formation at the insulator \cite{Tanaka1}.
The maximum Josephson current $I_{C}(T)$
is shown to have an anomalous temperature $(T)$
dependence, that is,
$I_{C}(T)$ becomes  proportional to $T^{-1}$ at low temperatures
when the ZES are formed at both interfaces \cite{Tanaka1,Barash,Riedel}.
\par
On the other hand,
the influences of the exchange interaction
on the Andreev reflection \cite{Andreev} have been
analyzed based on the Bogoliubov equation \cite{deJong}.
The theory has been extended to
$d_{{x}^{2}-{y}^{2}}$-wave and $p$-wave symmetries
recently, and the charge- and spin-transport
properties between ferromagnet and superconductors
have been revealed
\cite{Zutic,Kashiwaya99,Ting,Yoshida}.
In these theories, it has been shown
that the amplitude of the Andreev reflection
and the influence of the ZES are suppressed
with the increase of the exchange interaction
when the Cooper pairs are formed between
two electron with antiparallel spins.
This effect mainly originates from the breakdown of
the retro-reflectivity in the Andreev reflection process.
\par
At this stage, it is an interesting topic to clarify
how the above-mentioned properties affect the Josephson current.
This view point is really important,
because a recent progress in thin film technology
makes it possible to fabricate
hybrid structures containing both ferromagnets and
unconventional superconductors \cite{Goldman}.
In this paper, we present a formula
for the d.c. Josephson current in $S/F/S$ junctions
of which superconductors are in unconventional pairing states.
Although the formula can be applied for arbitrary symmetries,
we focus on the two cases, $i.e$. $d$-wave superconductor /
ferromagnet / $d$-wave superconductor ($d/F/d$) junctions
and $s$-wave superconductor / ferromagnet / $p$-wave superconductor
($s/F/p$) junctions.
\par
For the model of the calculation, we consider
two-dimensional $S/F/S$ junctions in the clean limit.
We assume a situation that
the ferromagnet, of which thickness is $d$, is inserted
between two semi-infinite superconductors (see Fig. 1).
The insulators located at the $F/S$ interfaces ($x=0$ and $x=d$)
are described by  potential
$V(\mbox{\boldmath $x$})$
\{$V(\mbox{\boldmath $x$}) = H[\delta(x) + \delta(x-d)]$,
where $\delta(x)$ $[\delta(x-d)]$
and $H$ are the $\delta$-function and its amplitude, respectively.
The effective mass
{\it m} is assumed to be equal
both in the ferromagnet and
in the superconductors.
For the model of the ferromagnet,
we adopt the Stoner model where
the effect of spin polarization is described by
the one-electron Hamiltonian with an
exchange interaction
similarly to the case of Ref. 15. 
For the description of the unconventional  superconductors,
we apply the quasi-classical approximation where the Fermi-energy
$E_{F}$ in the
superconductor is much larger than the pair potential amplitude
following the model by Bruder \cite{Bruder}.
The assumed  spatial dependence of the pair potential 
for the quasiparticle injected from the 
left superconductor with spin index $\sigma$ (see Fig. 1)
is described by
\begin{equation}
\label{5.5.0}
\Delta_{\sigma}(x,\theta)
=
\left\{
\begin{array}{ll}
\Delta_{L\sigma}(\theta)\exp(i\varphi_{L}), & (x<0) \\
\Delta_{R\sigma}(\theta)\exp(i\varphi_{R}), & (x>d)
\end{array}
\right. \
\end{equation}
\noindent
with $\Delta_{L(R)\sigma}(\theta)
=\Delta_{0}(T)f_{L(R)\sigma}(\theta)$,
where $f_{L(R)\sigma}(\theta)$
is a form factor  which specifies the
symmetry of the pair potentials.
The quantity $\theta$ and $\varphi_{L(R)}$
denotes an injection angle of the quasiparticles and
the external phase of the pair potential
of left [right] superconductor measured from
the normal to the interface, respectively.
The wave-vector of quasiparticles
in the ferromagnet for the spin-up [down] and
in the superconductor
are expressed as
$k_{N,\uparrow} =
\sqrt{\frac{2m}{\hbar^{2}}(E_{F} + U)}$
[$k_{N,\downarrow} = \sqrt{\frac{2m}{\hbar^{2}}
(E_{F} - U)}$], and
$k_{S} =\sqrt{\frac{2mE_{F}}{\hbar^{2}} }$
respectively \cite{deJong}.
The temperature dependence of $\Delta_{0}(T)$ is assumed to obey the
BCS relation with transition temperature $T_{C}$. \par
In the following, we  calculate the
Josephson current in $d/F/d$ junctions
where form factors are described by
$f_{L,\uparrow}(\theta)= -f_{L,\downarrow}(\theta)= 
\cos[2(\theta-\alpha)]$ and
$f_{R,\uparrow}(\theta)= 
-f_{R,\downarrow}(\theta)= 
\cos[2(\theta-\beta)]$.
The quantity $\alpha$ [$\beta$] is
the angle between the normal to the interface
and the crystal axis of the left [right] superconductors
[see  Fig. 1(b) of Ref. 11(b)].
Similarly to a previous theory \cite{Furusaki},
the Josephson current is described by
the coefficients of the Andreev reflection obtained
by solving the Bogoliubov equation.
To reduce the complexity,
we choose asymmetric junction ($\alpha=-\beta$) configurations.
After straightforward manipulations, the Josephson current is
expressed by
\begin{equation}
R_{N}I(\varphi)
=W
\int^{\pi/2}_{-\pi/2} d\theta
\sum_{\omega_{n}}
{\rm Real}
\left\{
\frac{ 4B_{2} \Gamma_{L} \tilde{\Gamma}_{L} \eta
\sin \varphi \cos \theta }
{(1 + \Gamma_{L}\tilde{\Gamma}_{L})^{2} B_{1}
+ C_{+} + C_{-}\Gamma_{L}^{2}\tilde{\Gamma}_{L}^{2}
+ 2B_{2}\Gamma_{L}\tilde{\Gamma}_{L}\cos(\varphi) \eta }
\right\}
\end{equation}
\[
B_{1}
=[(1-\lambda_{1}-iZ)(1+\lambda_{1}+iZ)
-(1+\lambda_{1}-iZ)(1-\lambda_{1}+iZ)\exp(2ip^{+}_{\uparrow}d)]
\]
\begin{equation}
\times [(1-\lambda_{2}+iZ)(1+\lambda_{2}-iZ)
-(1+\lambda_{2}+iZ)(1-\lambda_{2}-iZ)\exp(-2ip^{-}_{\downarrow}d)]
\end{equation}

\[
B_{2}=16 \lambda_{1} \lambda_{2}
\]

\[
C_{\pm}
=\pm 2(\lambda_{1} + \lambda_{2})
\{(1\pm \lambda_{1} \pm iZ)(1 \pm \lambda_{2} \mp iZ)
- (1 \mp \lambda_{1} \pm iZ)(1 \mp \lambda_{2} \mp iZ)
\exp[2i(p^{+}_{\uparrow}-p^{-}_{\downarrow})d] \}
\]
\[
\pm 2(\lambda_{1} - \lambda_{2})
\]
\begin{equation}
\times[(1 \mp \lambda_{1} \pm iZ)(1 \pm \lambda_{2} \mp iZ)
\exp(2ip^{+}_{\uparrow}d)
-(1 \pm \lambda_{1} \pm iZ)
(1 \mp \lambda_{2} \mp iZ)\exp(-2ip^{-}_{\downarrow}d)]
\end{equation}

\[
\Gamma_{L}=
\frac{\Delta_{L\sigma}(\theta_{+})}{\Omega_{L,+} + \omega_{n}},
\tilde{\Gamma}_{L}=\frac{\Delta_{L\sigma}(\theta_{-})}
{\Omega_{L,-} + \omega_{n}},
\Omega_{L,\pm}={\rm sgn}(\omega_{n})
\sqrt{\omega_{n}^{2} + \mid \Delta_{L\sigma}(\theta_{\pm}) \mid^{2} }
\]
\[
p_{\uparrow}^{+}=
\sqrt{\frac{2m}{\hbar^{2}}
(i\omega_{n} + E_{F}\cos^{2} \theta + U)},
p_{\downarrow}^{-}=
\sqrt{\frac{2m}{\hbar^{2}}
(-i\omega_{n} + E_{F}\cos^{2} \theta - U)},
\]
with
$\eta=\exp[i(p^{+}_{\uparrow}-p^{-}_{\downarrow})d]$,
$\theta_{+}=\theta$, $\theta_{-}=\pi-\theta$,
$W=\pi \bar{R}_{N} k_{B}T/e$,
$Z=Z_{0}/\cos\theta$, and $Z_{0}=2mH/(\hbar^{2}k_{S})$.
The normal resistance $\bar{R}_{N}$ of the junction is
given by
\begin{equation}
\bar{R}_{N}^{-1}
=\frac{1}{2}\sum_{j=1,2} \int^{\pi/2}_{-\pi/2}
\cos \theta (1 - \mid r_{j} \mid^{2}) d\theta
\end{equation}

\begin{equation}
r_{j}=
\frac{e^{2ik_{F}\cos\theta d \lambda_{j}}
[1-(\lambda_{j} - iZ)^{2}]-[1-(\lambda_{j} + iZ)^{2}]}
{e^{2ik_{F}\cos\theta d \lambda_{j}}(1 -\lambda_{j} + iZ)^{2}
-(1 + \lambda_{j} + iZ)^{2}]}
\end{equation}
for $j=1$ (up spin) and $j=2$ (down spin) with
$\lambda_{1}=\sqrt{ 1 + U/(E_{F}\cos^{2}\theta)}$
and
$\lambda_{2}=\sqrt{ 1 - U/(E_{F}\cos^{2}\theta)}$. \par
Similar to the results by the TK formula, the Josephson current
contains various $\theta$ components,
hence the properties of the junction is determined by the integration
on $\theta$.
In order to visualize the $\pi$-junction formation,
we define $I_{p}(T)=I(\varphi_{M})$,
where  the magnitude of the Josephson current
shows its maximum in $0<\varphi_{M}<\pi$.
Figure 2 shows $X$-dependence of $I_{p}(T)$ for
$d/F/d$ junction.
The sign of $I_{p}(T)$ changes from positive (negative) to
negative (positive) for curve $a$ ($b$).
These periodic changes between 0-junction and $\pi$-junction
are quite similar to those obtained
in the conventional
ones \cite{Buzdin1,Buzdin2}.
The origin of the periodic
change can  easily be checked analytically
for larger magnitude of $Z$.
In such a case, the Josephson current is rewritten as
\begin{equation}
R_{N}I(\varphi)
=W \int^{\pi/2}_{-\pi/2} d\theta
\sum_{\omega_{n}} {\rm Real}
\left\{
\frac{
16\lambda_{1}\lambda_{2} \Gamma_{L} \tilde{\Gamma}_{L}
\cos\theta \sin\varphi}
{Z^{4} \sin(p_{\uparrow}^{+}d) \sin(p_{\downarrow}^{-}d) 
(1 + \Gamma_{L}\tilde{\Gamma}_{L})^{2} }
\right\}
\end{equation}
For $X=0$, since
$p_{\downarrow}^{-}=p_{\uparrow}^{+ *}$
and $\lambda_{1}=\lambda_{2}$ is satisfied,
although an oscillatory
change of $I_{p}(T)$   due to the Friedel
oscillation exists,
$I_{p}(T)$ is always positive [negative]
for $\alpha=0$ (curve $a$ of Fig. 2) [$\alpha=\pi/4$ (curve $b$ of Fig. 2)].
However, for finite $X$,
$I_{p}(T)$ changes its sign thorough the sign-change of the factor
${\rm Real}\{ \lambda_{2}/
[\sin(p_{\uparrow}^{+}d) \sin(p_{\downarrow}^{-}d)] \}$.
On the other hand, as for the temperature dependence of the magnitude of $I_
{p}(T)$,
a rapid increment at low-temperature is obtained
for $\alpha \neq 0$ with larger $Z$.
This feature is
similar to those obtained
in the calculation of the TK formula, and originates from the ZES formed at
the interface.
The most interesting result, peculiar to the present model,
is obtained
for $\sin^{-1}(k_{N,\downarrow}/k_{S})<\theta<\pi/2$
when the Andreev reflection at the $S/F$ interface
becomes an evanescent wave [the virtual Andreev reflection
(VAR) process
described in Ref. 17 
as shown in  Fig. 1.
Since the ferromagnet behaves as if it were an insulator,
the Josephson current shows an exponential dependence
as a function of $d$.
Also the Josephson current is drastically suppressed
as  $X$ becomes larger.
This is  because the angle region
where the Josephson current is carried by the VAR process
monotonically increases with the increase of $X$.
Such a Fermi surface effect on the d.c. Josephson current has been
never discussed in previous theories.
\par
Next, let us move to an $s/F/p$ junction as a typical example  for the 
singlet superconductor / ferromagnet / triplet superconductor junction 
where 
$f_{R,\uparrow}(\theta_{+})= f_{R,\downarrow}(\theta_{+})=
\exp(-i\theta)$ and 
$f_{R,\uparrow}(\theta_{-})= f_{R,\downarrow}(\theta_{-})=
-\exp(i\theta)$ are satisfied.
Resulting Josephson current is given by
\begin{equation}
R_{N}I(\varphi)=
W
\sum_{\omega_{n}}
\int^{\pi/2}_{-\pi/2} d\theta
{\rm Imag}
\left\{
\frac{4B_{2}\Gamma_{L}\Gamma_{R} \eta \sin \varphi \cos\theta}
{[(1 + \Gamma_{L}^{2})(1- \Gamma_{R}^{2})B_{1}
+ C_{+}
- C_{-}\Gamma_{L}^{2}\Gamma_{R}^{2}]
+2iB_{2}\eta \Gamma_{L}\Gamma_{R} \sin \varphi  }
\right\}
\end{equation}
with
$\Gamma_{R}=
\frac{\Delta_{R\sigma}(\theta_{+})}{\Omega_{R,+} + \omega_{n}}$
$\Omega_{R,+}={\rm sgn}(\omega_{n})
\sqrt{\mid \Delta_{R\sigma}(\theta_{+}) \mid^{2} + \omega_{n}^{2}}$, 
$f_{L,\uparrow}(\theta_{+})=-f_{L,\downarrow}(\theta_{+})=1$. 
Only the lowest order coupling remains finite
at the limit of large magnitude of $Z$,
then $R_{N}I(\varphi)$ converges to
\begin{equation}
R_{N}I(\varphi)
=W \int^{\pi/2}_{-\pi/2} d\theta
\sum_{\omega_{n}}
\frac{ 16\lambda_{1} \cos^{2}\theta \sin\varphi
\Gamma_{L} (1 -\mid \Gamma_{R} \mid^{2})}
{(1+\Gamma_{L}^{2}) \mid 1 - \Gamma_{R}^{2} \mid^{2}Z^{4}}
{\rm Imag}
\left\{
\frac{\lambda_{2} }
{\sin(p_{\uparrow}^{+}d) \sin(p_{\downarrow}^{-}d)}
\right\}
\end{equation}
When $X=0$, $s/F/p$ junction
reduces to an usual Josephson junction
formed between $s$-wave and $p$-wave superconductors.
In this case, as shown in the curve $a$ of Fig. 3,
$I(\varphi)$ has a period of $\pi$
(not $2\pi$),
since the lowest order Josephson coupling
between the two superconductor diminishes due to the
difference of the parity and the rotational symmetry,
$i.e.$, $SU(2)$ symmetry, in the spin space.
As seen in Eq. (9),  since $\lambda_{2}=1$ and
$p_{\uparrow}^{+}=p_{\downarrow}^{-*}$ is satisfied for $X=0$,
$R_{N}I(\varphi)$ vanishes for larger $Z$.
However, for finite $X$,
${\rm Imag}\{\lambda_{2}/[\sin(p_{\uparrow}^{+}d)\sin(p_{\downarrow}^{-*}d)]\}$
is non-zero as shown in Eq. (9).
Then the cancellation between up spin injection and down spin injection
becomes incomplete and the imaginary part of Eq. (9)
also becomes finite.
As the result, the cancellation in the lowest order
is lifted and the component with period $2\pi$ recovers
as shown in curves $b$ and $c$ in Fig. 3.
The recovery in the lowest order enhances the magnitude of the
d.c. Josephson current once.
As shown in curve $d$ in Fig. 3, $I_{p}(T)$
is reduced
with the further increment of $X$.
This reduction is caused by the
breakdown of the retroreflectivity in the Andreev reflection
similar to the cases of $d/F/d$.
\par
In the above, the temperature dependence of
$X$ has not been taken into account.
If the transition temperature of the ferromagnet $T_{\theta}$
is sufficiently larger than $T_{C}$, $X$  can be regarded as a
constant for all temperatures below  $T_{C}$.
However, when $T_{\theta} \sim T_{C}$, $I_{p}(T)$ is expected to
have a non-monotonous temperature dependence due to
the competition between the pair potential amplitude enhancement 
(positive contribution) 
and the polarization  enhancement (negative contribution) 
\cite{Buzdin1}. 
For simplicity, consider a $d/f/d$ junction 
and assume the temperature dependence of 
$X(T)$ as $X(T)=X_{0}\sqrt{(1 - T/T_{\theta})}$. 
As shown in Fig. 4, $I_{p}(T)$ changes drastically 
from $0$-junction ($\alpha=0$ curve $a$)
to $\pi$-junction ($\alpha=\pi/4$ curve $b$) at
$T=T_{\theta}$. Such an anomalous feature
may be accessible in actual experiment. \par
In this paper, we present a formula of the Josephson current
in $S/F/S$ junctions where the Cooper pair is formed between two
electrons with antiparallel spins.
In most cases,
the break down of the retro-reflectivity
of the Andreev reflection reduces the magnitude of the
Andreev reflection and the resulting
Josephson current is suppressed.
However, in the case of Josephson junctions formed between
two superconductors with different parities,
the Josephson current can be enhanced
by the exchange interaction due to the
recovery of the first order Josephson
coupling.
Throughout this paper,
the spin-orbit scattering is ignored.
The essential conclusion may not be changed even if
we take into account of this effect if the magnitude of
the spin orbit coupling is small,
because the above feature is governed by the parity.
>From an experimental view point,
it is possible to make a
Josephson junction with $d$-wave superconductor and
ferromagnet using
Mn oxides compound and
high-$T_{c}$ superconductors \cite{Goldman}
or  Sr$_{2}$RuO$_{4}$.
We hope anomalous properties predicted in this paper
will be detected near future. \par
\vspace{1.0cm}
We would like to thank N. Yoshida,
J. Inoue, M. Koyanagi, and M. R. Beasley for fruitful discussions.
This work has been partially supported by the
Core Research for Evolutional
Science and Technology (CREST) of the Japan
Science and Technology
Corporation (JST) and a Grant in aid for Scientific Research
from the Ministry of Education, Science, Sports and Culture. \par

\newpage
\noindent
Figure Captions \par
\noindent
Fig. 1 A schematic illustration of the reflection and transmission
process at the interface of $S/F/S$ junction. \par
\noindent
\noindent
Fig. 2 $X$ dependence of $I_{p}(T)$ for
$d/F/d$ junction for $dk_{S}=5$ and $Z_{0}=5$ at
$T=0.1T_{d}$. a: $\alpha=0$ b:$\alpha=\pi/4$. \par
\noindent
Fig. 3 Current phase relation $I(\varphi)$ for $s/f/p$ junction
with $dk_{S}=1$ and $Z_{0}=5$ at sufficiently  low temperatures.
a: $X=0$ b: $X=0.1$ c: $X=0.5$ and $X=0.99$.
\noindent
Fig. 4 Temperature dependence of $I_{p}(T)$ for $d/f/d$ junction
with $Z_{0}=5$, $dk_{S}=5$, $X_{0}=0.5$,
and $T_{\theta}=0.5T_{C}$.
a: $\alpha=0$ b: $\alpha=0.25\pi$.
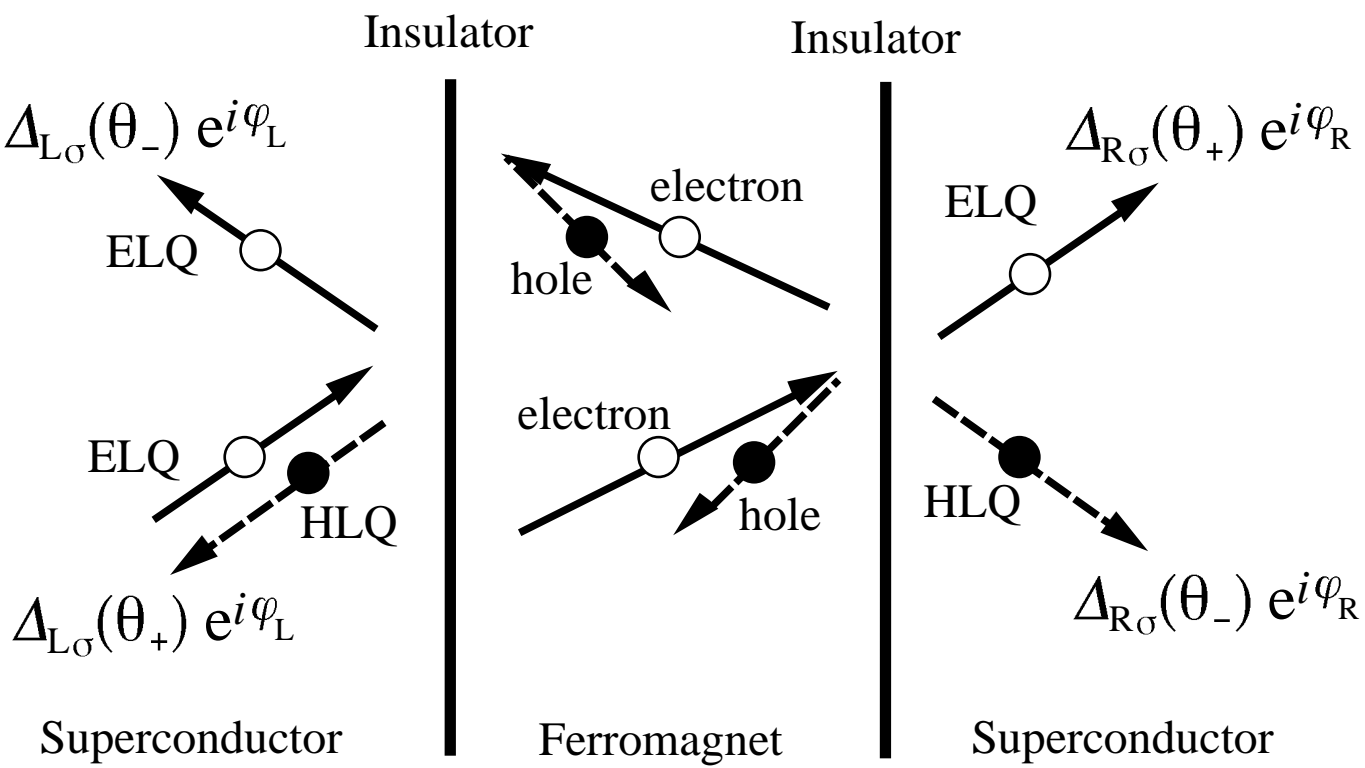
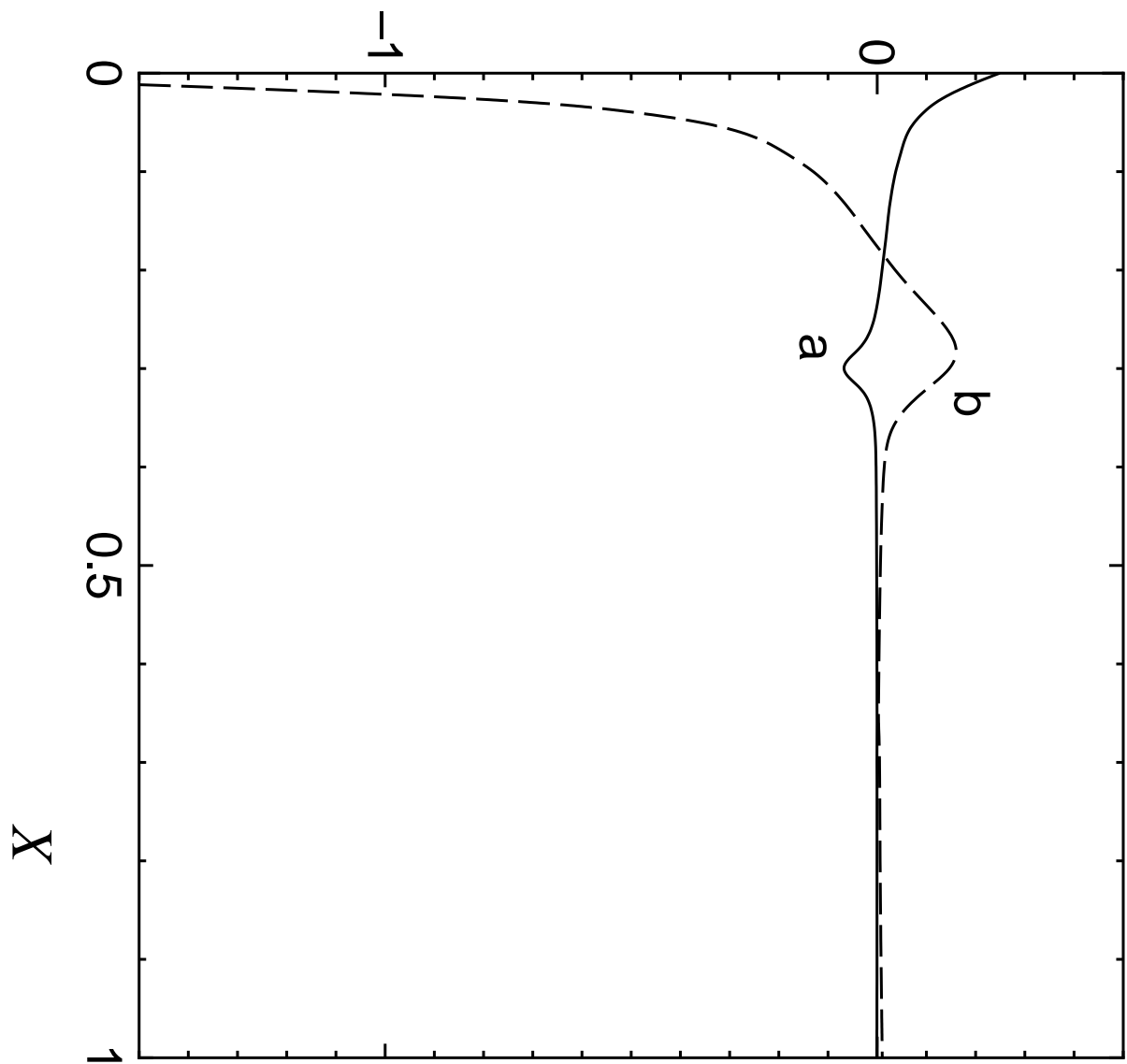
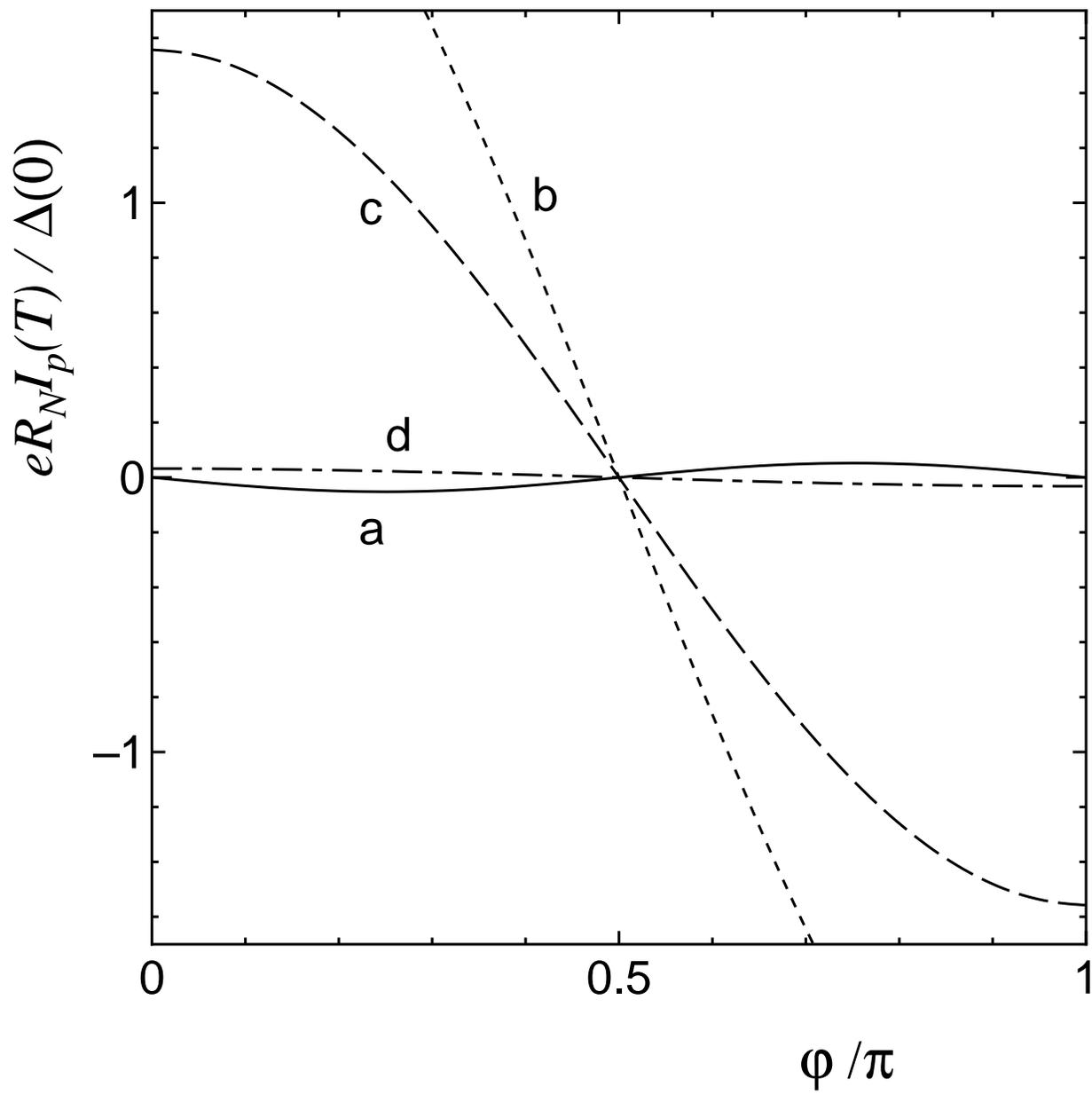
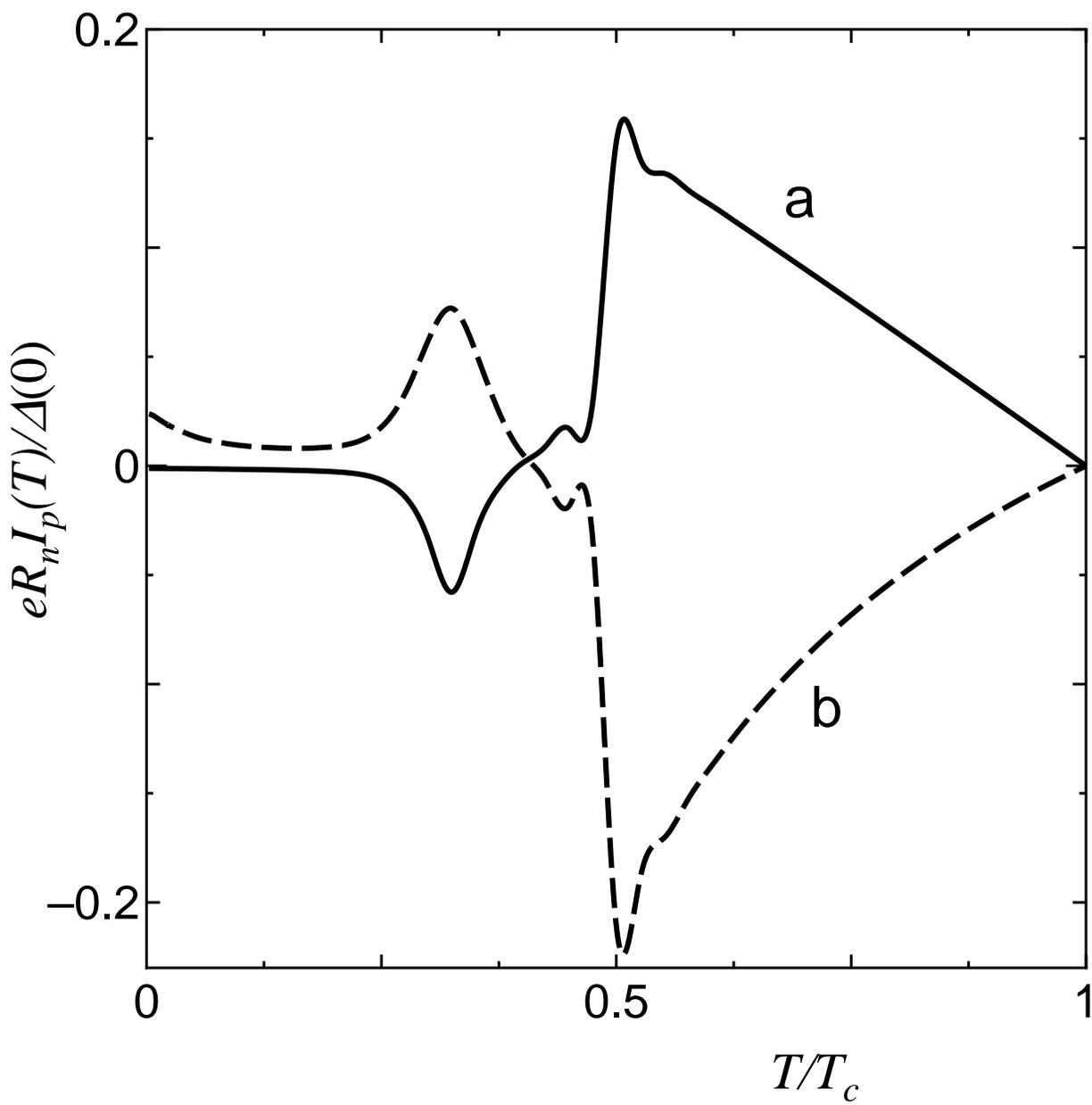
\end{document}